\documentclass[fleqn,10pt]{wlscirep}

\usepackage[utf8]{inputenc}
\usepackage[T1]{fontenc}

\usepackage{color}
\usepackage{comment}

\title{Urbanization and Economic Complexity}

\author[1*,2]{Riccardo Di Clemente}
\author[3]{Emanuele Strano}
\author[2]{Michael Batty}

\affil[1]{Department of Computer Sciences, University of Exeter, Exeter, EX4 4QF, UK.}
\affil[2]{The Centre for Advanced Spatial Analysis, University College London, London WC1E 6BT, UK.}
\affil[3]{MindEarth, 2502 Biel/Bienne, Switzerland}

\affil[*]{Corresponding author: r.di-clemente@exeter.ac.uk}

\begin{abstract}
Urbanization plays a crucial role in the economic development of every country. The mutual relationship between the urbanization of any country and its economic productive structure  is far from being understood. We analyzed the historical evolution of product exports for all countries using the World Trade Web (WTW) with respect to patterns of urbanization from 1995-2010. Using the evolving framework of economic complexity, we reveal that a country's economic development in terms of its production and export of goods, is interwoven with the urbanization process during the early stages of its economic development and growth. Meanwhile in urbanized countries, the reciprocal relation between economic growth and urbanization fades away with respect to its later stages, becoming negligible for countries highly dependent on the export of resources where urbanization is not linked to any structural economic transformation.
\end{abstract}
\begin{document}

\flushbottom
\maketitle
\thispagestyle{empty}

\section*{Introduction}

It is an established fact that urbanization in developed countries is accompanied by economic growth and industrialization which mutually self-reinforce one another\cite{Martin_2001}.
This historic pattern generates an expectation of a virtuous circle between economic growth and urbanization regardless of local conditions\cite{buckley_2008,Duranton_2014}.
From classic urban economic theories\cite{jacobs2016economy,Glaeser_1992} to the more recent scaling approach to cities\cite{Bettencourt:2007aa,Youn_2016}, the growth of urban population has routinely been used as a proxy for economic growth. This pattern has also been observed in rapidly developing countries such as China and India but it cannot be considered a universal blueprint\cite{Ravallion_2007} for deviations from this norm have not yet been fully explained.

In fact, as pointed out in several studies\cite{Montgo_2008,PETRAKOS19891269,Gollin_2015,Glaeser_2014}, the increasing urbanization rate in persistently poor and non-industrialized countries poses an important dilemma for urban economic theory. Why, given the same rate of urbanization, does Asia contain a number of explosive economies while sub-Saharan Africa has seen very little growth? Moreover, in developed and advanced industrialized economies, is there appears to be a competitive advantage in continuing this urbanization process indefinitely?

There are several theories aimed at explaining urbanization processes. Some argue that rural poverty moves people to cities as was clearly the case in 19th century Europe and America\cite{Ades_1995}, driving the transformation from an agricultural to an industrial-service based economy\cite{DAVIS200398,Henderson_2003}. 
Others argue that in the last decades there has been urban-biased public policy that has led to over-urbanization\cite{Glaeser_2014}.

The most intriguing approach however is rooted in the mutual indirect effects of the World Trade Web (WTW) on global urbanization\cite{Fujita_2003,krugman1993geography}. The dominant idea is that in open economies, domestic communities (such as cities) can trade easily with other communities, boosting their exports in substituting industrialization for urbanization policy\cite{KRUGMAN1996137}. In simple terms, the commodities can flow more freely using urban agglomerates as nodes in the trading networks between countries, generating the ever present virtuous circle between economic growth and urbanization.

Starting from this theory, we analyze the WTW to explore the mutual relationship between the urbanization of the countries and their economic production structure using the Economic Complexity (EC) framework. Economic Complexity, \cite{Hidalgo:2009aa,tacchella2012new,cristelli2013measuring,Hidalgo:2007aa,tacchella2018dynamical,tacchella2013economic,albeaik2017improving,e20100783,Sciarra2020} is a new and expanding field in the economic analysis, which proposes "Fitness" and "Complexity" metrics to quantify the fitness or competitiveness of countries and the complexity of products from a country's basket of exports. The main focus of EC is based on a bipartite representation of the World Trade Web where the nodes represent the set of world-countries and the set of exported products defined as different entities. Countries and products are connected to one another by imposing a threshold based on their Revealed Comparative Advantage (RCA)\cite{doi:10.1111/j.1467-9957.1965.tb00050.x} which defines the criterion for the existence of  relations.

The Fitness and Complexity algorithm is a kind of PageRank method applied to WTW, where Fitness $F_c$ is the quantity for country $c$, and Complexity $Q_p$ is the quantity for products $p$. The idea at the basis of the algorithm is that the countries with the highest fitness are those which are able to export the highest number of the most exclusive products i.e. those with the highest complexity. On the other hand this complexity is non linearly related to the fitness of its exporters so that products exported by low fitness countries have a low level of complexity and high complexity products are exported by high fitness countries only.

The Fitness metric is valuable in quantifying a country's productive structure and structural transformations which enable one to predict its future economic growth\cite{tacchella2018dynamical}. It correlates with the extent of economic equality\cite{hartmann2017linking} and it has been used to analyze  the country's growth path to industrialization \cite{pugliese2017complex}.

In this work, we used a data driven approach borrowing tools recently introduce by statistical physics and network science to improve our understanding of the complex dynamics of human societies, with the aim of finding innovative insight\cite{perc2019social} to link urbanization process with the evolution of the international trade.

We couple the WTW data with the urbanization level of more than 146 countries worldwide, and analyze this between 1995-2010 thus capturing the fingerprint of urbanization on countries' productive systems through the lens of their exports. We notice that in rural economies, the increase in urban population fosters structural changes in industrial exports. It boosts the country's diversification improving the country fitness, and allowing the export of more complex products. These economic transformations fade away in countries that already have a high level of urban population (more than $60\%$) where there is no relation between the urbanization process and the country's fitness.

Within the sub-Saharan countries, we capture those where the virtuous circle between economic growth and urbanization is fostering structural changes in those countries' productive systems. On the other hand within countries with economies based on raw materials, we assess the implementation of policy leading to urbanization that does not support any structural transformations of their basket of exports.

\vskip10pt

\section*{Results}

\subsection*{Economic Complexity and Urbanization}

We represent the WTW as a bipartite network, i.e. by considering the set of world-countries and the set of products as different entities and linking a given country to a given product if (and only if) the former exports to the latter above a certain threshold (the so-called Revealed Comparative Advantage - RCA)\cite{doi:10.1111/j.1467-9957.1965.tb00050.x}. RCA is a general criterion adopted in order to understand whether a country can be considered, or not, a producer of a particular product. It quantifies how much the export of a given product $p$ is relevant for the economy of a country $c$ in relation to the global export of $p$ for all countries (See Methods Section below).

The country's fitness and product's complexity are the result of a non-linear iterative map applied to the WTW matrix $M$ \cite{Hidalgo:2009aa,caldarelli2012network,tacchella2012new} (See Methods below).

Through the algorithm's iterations, products exported by low fitness countries have a low level of complexity while high complexity products are exported by high fitness countries only. The countries' composition of their export basket depends on their fitness. Fitness and Complexity are thus non-monetary indicators of the economy's development: the fitness represents a measure of tangible and intangible assets and capabilities, which drive the country's development, such as political organization, its history, geography, technology, services, and infrastructures\cite{cristelli2013measuring}. Meanwhile complexity measures the necessary capabilities which must be owned by a country in order that it can produce and then export the resulting product.

\begin{figure}[!hbt]
\centering\includegraphics[width=0.99\textwidth]{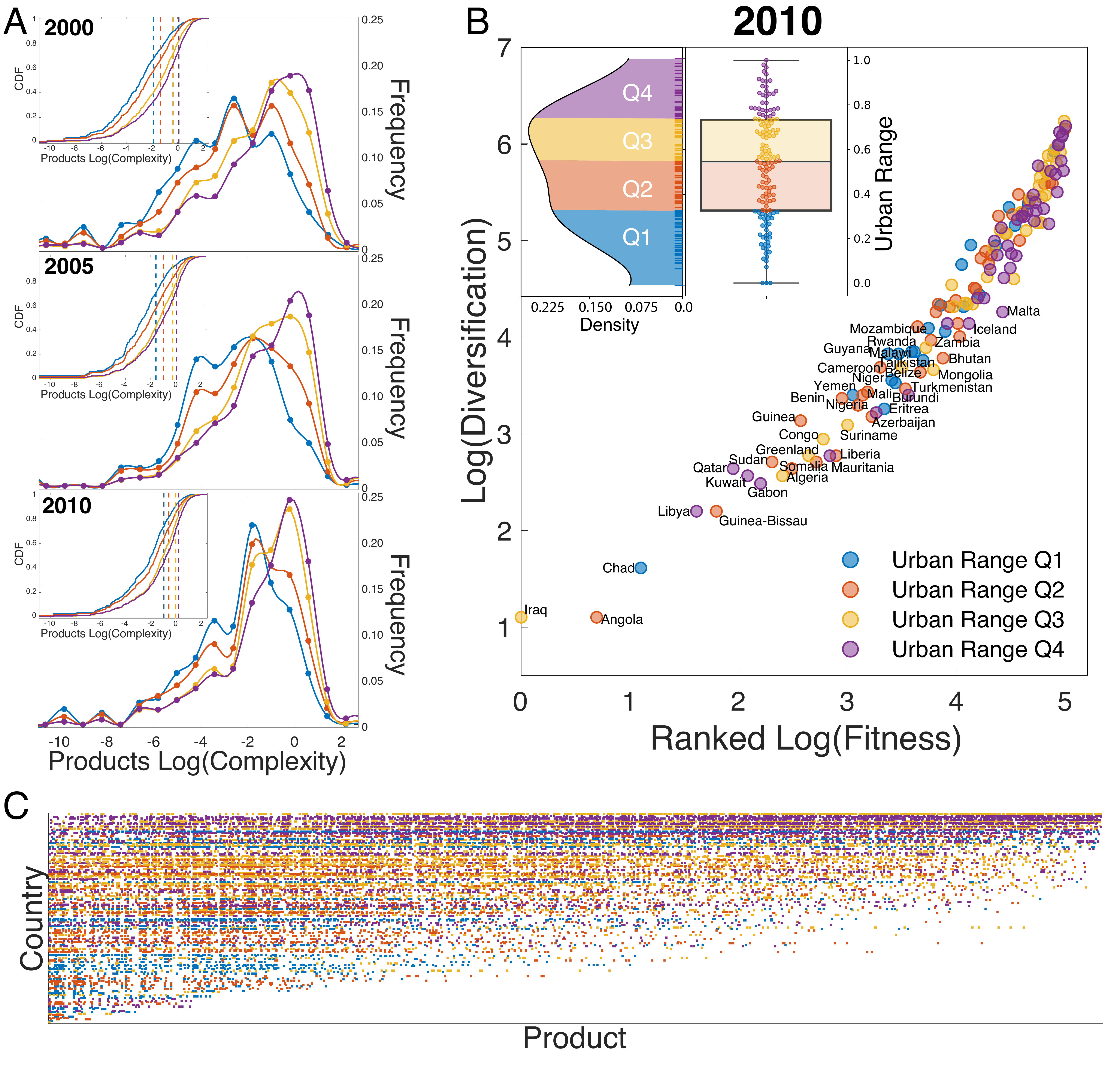}
\caption{\textbf{A.} Distribution of exported products complexity by different urbanization levels through the 2000-2010. There is a shift of lower urbanized countries towards the export of more complex products \textbf{B. caption} Distribution of the Urban Range (percentage of the total population living in urban areas) of the $146$ countries analyzed. \textbf{B.} Ranked country fitness vs products export diversification, the highly diversified countries are one's with more fitness and high urbanization, meanwhile low urbanized country are in the center bottom of the scatter plot, with some exceptions such as those with links to the oil countries. \textbf{C.}  Matrix of the country exports in 2010, reordering the countries and products by fitness and complexity; the color dots represent an exported product under the RCA threshold Eq.\ref{eq_rca}, the color gradient follows the urban range definition.}
\label{fig_1}
\end{figure}

Within this framework, we include the dimension of a country's degree of urbanization defined as the percentage of the total population living in urban areas. Our aim is to quantify the link between a country's urbanization process and their exports as a proxy for their industrial economic system. To disentangle the relation between country productivity systems and their urbanization, we have divided the set of countries in terms of their degree of urbanization, defined by the Urban Range, which is expressed in four quantiles $[Q1,Q2,Q3,Q4]$ (see urban range division in Fig.\ref{fig_1}B top).

More urbanized countries [Q3,Q4] in the early 2000s, export a wide range of complex products such as: textiles, heavy manufacturing industries, and IT while rural countries [Q1,Q2] export products that require a low level of sophistication such as raw materials and agricultural products (Fig.\ref{fig_1}A).

Highly urbanized countries maintain a similar distribution across the analysis years, with a long tail of low complexity products and a consistent increase in the number of high complexity products. On the other hand starting from 2005, we have noticed that rural countries change their export basket towards higher complexity products. This shift is shown by the cumulative distribution functions of the different Urban ranges that decrease their distance from one another over time  (see Fig.\ref{fig_1}A inset) together with their median and peak distance.

We notice that countries within the higher quantile of the Urban Range, Fig.\ref{fig_1}B, are the ones with higher fitness and higher diversification, whilst low urbanized countries have a low diversification and fitness.
Notable exceptions are countries with exports based on raw materials (i.e. Qatar, Kuwait, Gabon, Iraq, Libya). These countries reached higher levels of urbanization as result of policy decisions \cite{10.1093/jeg/1.1.81} meanwhile their exports are limited to a few products with low complexity.

The representation of the WTW in Fig. \ref{fig_1}C shows country exports in 2010 rearranged by ranked fitness and complexity. The country exports' diversification is related to the urbanization level. Low urbanized countries are at the bottom of the matrix with low fitness and lower degree of diversification, whilst the urbanized countries, with the most advanced economies lie at the top, with a high degree of product diversification with different levels of complexity and high fitness.

\subsection*{Exports Diversification and Urbanization}

It is known that low fitness countries have similar economies with low degrees of diversification and high similarity with respect to their export baskets\cite{saracco2015randomizing,saracco2016detecting,tacchella2012new} i. e. they produce and export few of the same low technology products. We captured a shift in the distribution of the exported products within the rural countries (Fig.1A). In particular, we noticed that rural countries start to produce and export more sophisticated products. This productive systems transformation in the EC literature is related to the development of new capabilities\cite{saracco2015innovation,tacchella2016build,Hidalgo:2007aa}.

Some questions from this analysis emerge: do the rural countries evolve their productive systems in the same way? and do they continue to produce and export the same products? Is the pattern of economic development entangled with urbanization in same fashion for each rural country?

\begin{figure}[!htb]
\centering\includegraphics[width=0.99\textwidth]{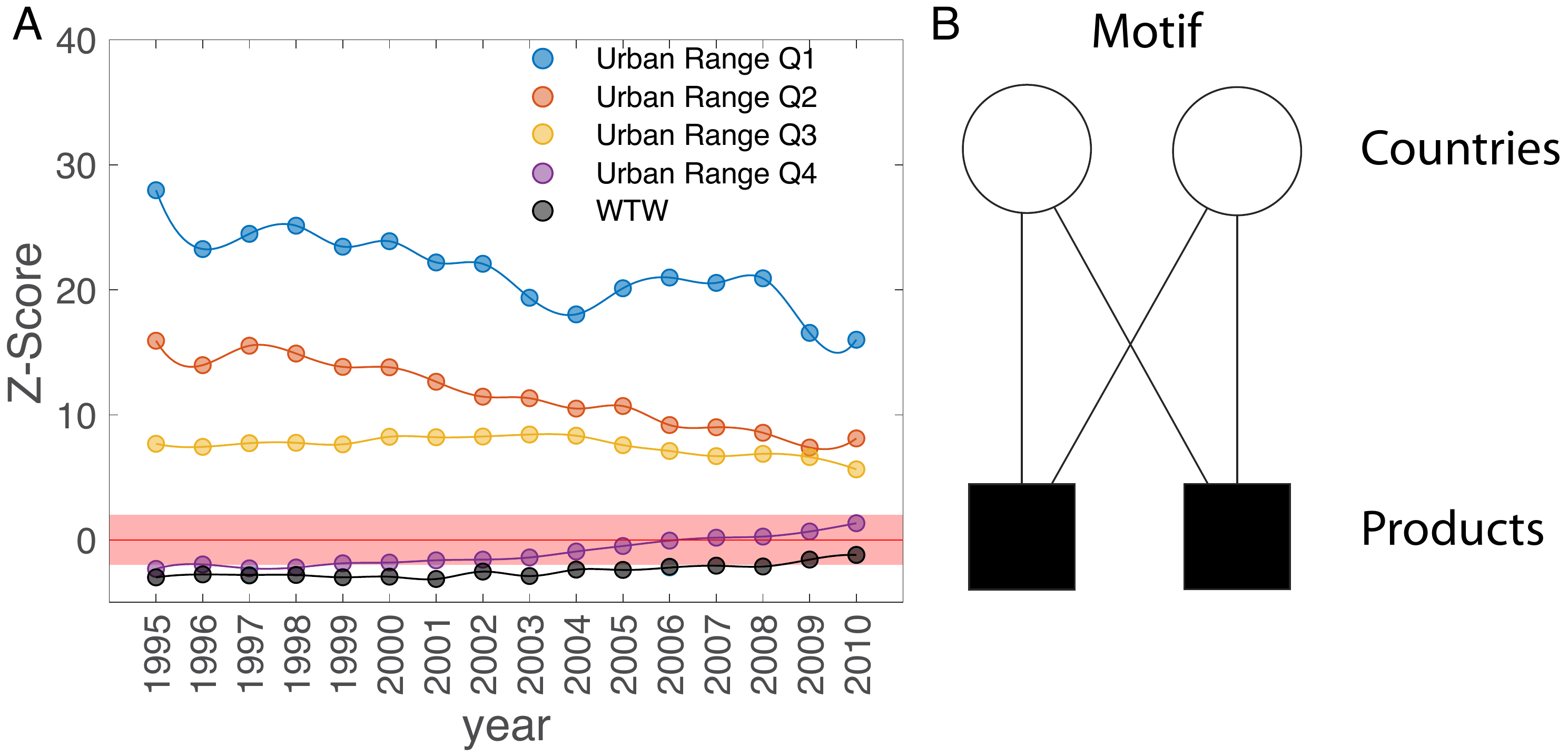}
\caption{\textbf{A} Z-score of the export similarity motif by country groups with different Urban Ranges and the Z-score of the whole WTW in black. \textbf{B} Similarity motif as the co-occurrence of any two countries as producers of the same couple of products \cite{saracco2016detecting,simmons2019bmotif}.}
\label{fig_2}
\end{figure}

We can measure indirectly the transformation of the productive systems by analyzing the evolution of WTW topology\cite{VanLidthdeJeude2019}.  In particular we can assess the changes of countries' similarities in their exports studying the abundance evolution of network motifs\cite{Gualdi2016}. A network motif is a particular pattern of interconnections occurring between the nodes of the network (i.e. between the countries and their products).
In our case we are interested in the abundance of the similarity motif $\mu_{sim}$ in Fig.\ref{fig_2}B (motifs 6\cite{simmons2019bmotif}, or X motif\cite{saracco2016detecting}): it quantifies the co-occurrence of any two countries as producers of the same couple of products as Eq.\ref{eq_sim} (and, viceversa, the co-occurrence of any two products in the basket of the same couple of countries). This represents the simplest motifs\cite{simmons2019bmotif} that  can quantify the similarities in the export countries' diversification which maintains a pairwise correlation within the products exported. Two economies with a fixed number of products exported are diversifying if the values of $\mu_{sim}$ is decreasing while their production similarity increases with high values of  $\mu_{sim}$.

To provide a benchmark and asses the $\mu_{sim}$ statistical significance of the WTW we use the Bipartite Configuration Model (BiCM)\cite{saracco2015randomizing} as a null model. This framework is valuable in the  analysis of the abundance of the bipartite motifs\cite{simmons2019bmotif},  enabling us to detect financial crisis effects on a country's export basket \cite{saracco2016detecting} as well as export similarities between countries with same level of economic development \cite{Saracco_2017}.

We generated $1000$ matrices using the BiCM\cite{saracco2015randomizing} (see Methods Section below) and we compare the observed abundances of the similarity motif (Eq.\ref{eq_sim}) in the real network with the corresponding expected values in the null ensemble using the Z-score.

The whole WTW manifests a progressive increase of the abundance of similarity motifs with respect to the null case\cite{saracco2016detecting} (Black line Fig.\ref{fig_2}A). Highly urbanized countries show a similar trend of increasing similarity in their products exports. This measure implies that rural economies are very similar with a higher abundance of the similarity motif with respect to the random case having a high value Z-score. Interestingly, low urban range countries diversifying between each other manifest an opposite trend. The exports diversification trends of the low urbanized countries coupled with the increasing complexity of the product exported imply a nontrivial connection between urbanization and production capabilities. This measure outlines how rural economies follow different development patterns based on their production systems. The urbanization phenomenon coupled with the capabilities already presented in the country enable the production of different sophisticated products depending on their environment.

\subsection*{Urbanization Growth and Country Fitness}

\begin{figure}[!htb]
\centering\includegraphics[width=0.99\textwidth]{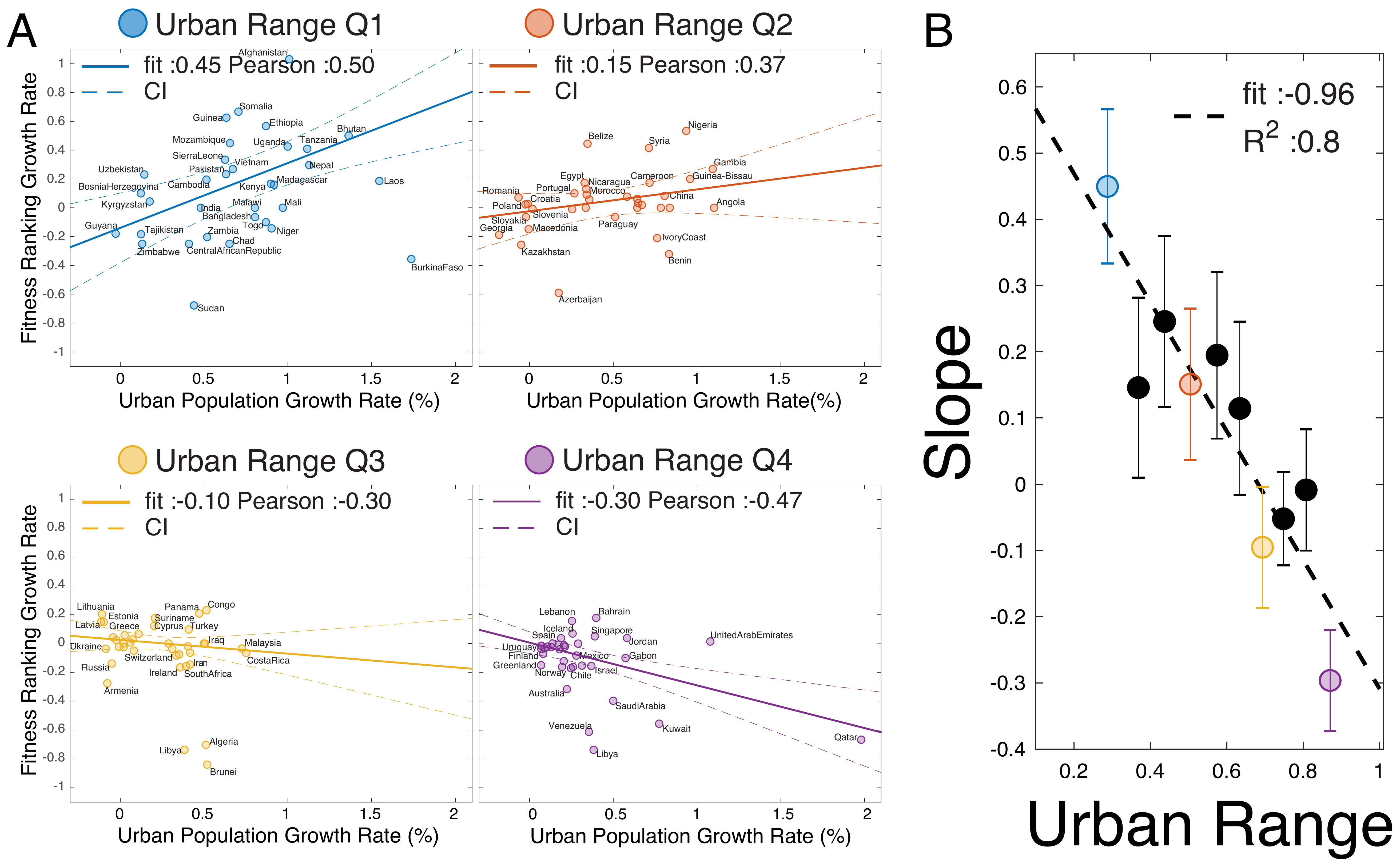}

\caption{\textbf{A} The Fitness Ranking Growth Rate vs. Urbanization Growth Rate. The effect of urbanization growth on the transformation of the economic systems (or vice-versa) is more relevant in low urbanize countries. \textbf{B} Slope coefficient of a sliding window across $25\%$ of the countries (corresponding to 36 countries) of its Fitness Ranking Growth Rate vs. Urban Population Growth Rate. The error bar corresponds to the fit's  $95\%$ confidence interval. The colors follow the Urban Range Scheme. }
\label{fig_3}
\end{figure}

The economic transformation of a rural country has an impact on its overall fitness value, and the competitiveness of its productive system. 
In this respect, the urbanization process is key element in a country's development and its economic growth\cite{BERTINELLI200480,10.1093/jeg/1.1.81}.
To assess the relation between the country's fitness and the urbanization process we analyzed the Urban Range growth rate in relation to the growth rate of country fitness ranking between 1995 and 2010, as we show in Fig.\ref{fig_3}.
The country fitness ranking is the country's ordered position with respect to the country's fitness value in a given year. The growth rate of the country's fitness ranking is an easily understood tool to compare the transformations of a country's productive systems with respect to its competitors. It has been proven a reliable tool in quantifying the country's relative degree of competitiveness across different years providing a more stable measurement than the raw fitness value \cite{Pugliese_2016}.

For each of the four Urban Range quantiles we find a linear relation between the urbanization rate and the Fitness ranking growth rate in Fig.\ref{fig_3}b. Increasing urbanization within lowly urbanized countries is interwoven with increasing Fitness. Meanwhile, the effects are minimal in highly urbanized countries (Urban Range Q3,Q4). We validate the urbanization/fitness relation analyzing a $25\%$ quantile sliding window on the whole urbanization distribution, which we show in black in Fig.\ref{fig_3}B.

We notice that in many rural economies, the urbanization process affects or has been affected by structural changes in its economic production. (An example are countries such as Uganda, Nepal, Somalia.) 
On other hand, there are many countries (such as IvoryCoast, Paraguay, Chad) where the urbanization process does not provide improvement in the fitness ranking \cite{remi2015Demography}.

The self-reinforced mechanism between urbanization and fitness reaches a plateau within the urbanized countries (Q3,Q4), where the urbanization does not affect or has not been affected by changes in fitness ranking. In this respect, the resource exports countries manifest a shift toward a negative relation between urbanization and fitness. In fact in countries that are heavily dependent on resource exports, urbanization appears to be concentrated in the cities where the economies consist primarily of non-tradeable services\cite{gollin2016urbanization}.
To support our result we provide the same analysis using instead of the Fitness Ranking metric, the Fitness, GPD and GDP Ranking respectively (see Methods section: Urban Range vs Fitness and GDP\ref{Urban}). We do not find any evidence of relation between the other three metrics and the urbanization rate.

\subsection*{Urban Fitness Trends}

The process of urbanization is often entangled with a country's industrialization \cite{Gollin_2015}. As countries develop, people move out of rural areas and agricultural activities into urban centers,  where they engage in manufacturing products \cite{10.1093/qje/qjs003} which are more sophisticated with higher complexity. 
This transformation is outlined by the increasing level of fitness of low urbanized countries that are involved in the urbanization process. To leverage this information and capture its trends, we define the country Urban Fitness $F_{c}^{\mbox{urb}}(t)=F_c(t)*U_c(t)$; 
this is the value of country fitness $F_c$ weighted by the percentage of urban population $U_c$. 

\begin{figure}[!htb]
\centering\includegraphics[width=0.99\textwidth]{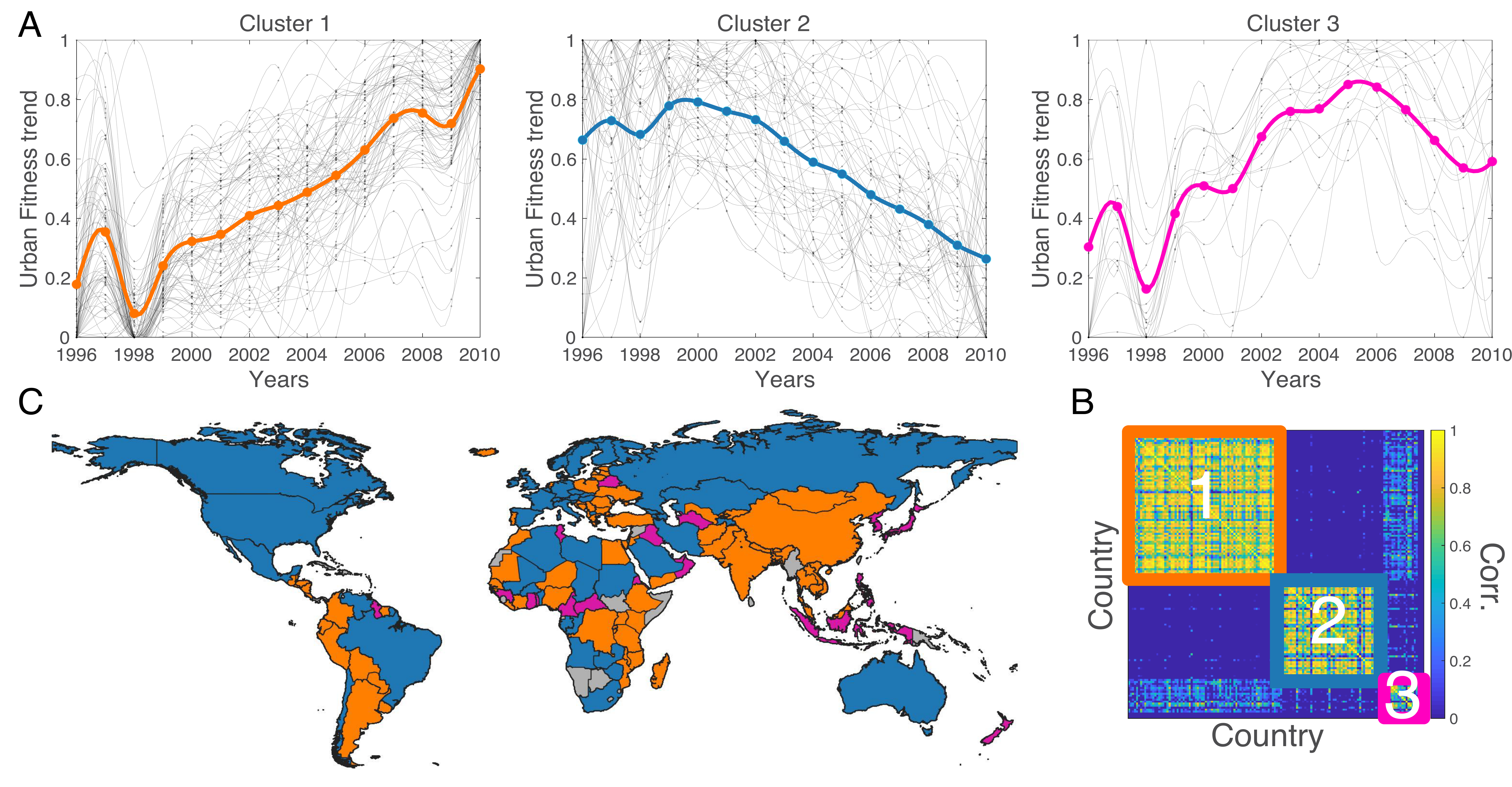}
\caption{\textbf{A.} Clusters of normalized Urban Fitness Trends. \textbf{B} Correlation Matrix of the countries urban fitness trends clustered with the Louvain algorithm. \textbf{C} Geographical cluster distribution. The map in this figure was created using the software QGIS.}
\label{fig_4}
\end{figure}

We cluster the countries Urban Fitness trends using the Louvain algorithm \cite{Blondel_2008} which is based on their correlation matrix shown in   Fig.\ref{fig_4}B. Three clusters emerge with high correlations disentangling the non-trivial geographical relations we show in Figs.\ref{fig_4}A-C.

In Figure \ref{fig_4}A countries with a clear urbanization trend (in orange) are ones with a stable increase in fitness ranking. Meanwhile the blue cluster contains developed countries, where the urbanization does not provide any new input to the economic
development and resource dependent countries, where the urbanization is not only lead by deep structural economic change. These results are in agreement with the Urban Range study in Fig. \ref{fig_3} that show a poor effect of the urbanization on the country fitness, implying that over a given value of urbanization, other factors have a more important role in economic development and growth. Finally, the third cluster (in red) are the countries without any clear trend and are thus uncategorized.

\section*{Discussion}

It is well-known that urbanization provides several advantages to the economics of scale and division of labour, boosting productivity and competition. It helps in accessing the labor force and inputting materials to the production process as well as decreasing the geographical distance between firms, reducing transaction costs, and fostering competition\cite{Turok_2013}. 
These urbanization advantages \cite{10.1093/icc/12.3.577} together with the appropriate bureaucratic environment \cite{10.1093/jeg/1.1.81}, investment in infrastructures \cite{ESFAHANI2003443} and companies market structure \cite{Di_Clemente_2014}, are some of intangible attributes, the capabilities, that a country needs to drive economic growth and innovation \cite{saracco2015innovation}.
We noticed that the country Fitness, the production and export of goods, is interwoven within the urbanization process during the early stages of country's economic development and growth. We show that the information carried by WTW can provide a different perspective on analyzing the complex process of urbanization, enlightening the relation between a country's exports, economic development and its urban growth.

\section*{Methods}
\subsection*{Data}

\subsubsection*{World Trade Web}

The dataset used in this work is the BACI (Base pour l'Analyse du Commerce International)  World Trade Database (Gaulier, S. Baci: International trade database at the product-level \href{http://www.cepii.fr/CEPII/fr/publications/wp/abstract.asp?NoDoc=2726}{http://www.cepii.fr/CEPII/fr/publications/wp/abstract.asp?NoDoc=2726} Date of access: 18/01/2021).
The data contains information on the trade of 200 different countries for more than 5000 different products, categorized according to the 6-digit code of the Harmonised System 2007 (\href{http://www.wcoomd.org/}{http://www.wcoomd.org/} Date of access: 18/01/2021.).
The products' sectors follows the UN categorization (\href{http://unstats.un.org/unsd/cr/registry/regcst.asp?Cl=8}{http://unstats.un.org/unsd/cr/registry/regcst.asp?Cl=8} Date of access: 18/01/2021).
We create a map between the two systems converting the HS2007 in to the ISIC revision 2 code at 2-digit (\href{http://www.macalester.edu/research/economics/PAGE/HAVEMAN/Trade.Resources/TradeConcordances.html\#FromISIC}{http://www.macalester.edu/research/economics/PAGE/HAVEMAN/Trade.Resources/TradeConcordances.html\#FromISIC} Date of access: 18/01/2021).
We represent the trade relation between the $145$ countries $c\in[1,C]$ and the $1131$ products $p\in[1,P]$ between the years $[1995,2010]$ throught the bipartite matrix $\tilde{M}$ with dimension $(C\times P)$ where each entry $\tilde{m}_{c,p}$ measures the export in US dollars.
The framework of the Economic-Complexity\cite{Hidalgo:2009aa,tacchella2012new,cristelli2013measuring,Hidalgo:2007aa} based on the interaction between countries and products is expressed by the application of the Revealed Comparative Advantage (RCA) \cite{doi:10.1111/j.1467-9957.1965.tb00050.x} threshold over the entries $\tilde{m}_{c,p}$:

\begin{equation}
	\mbox{RCA}_{c,p}=\frac{\frac{\tilde{m}_{cp}}{\sum_{1,P}^{p'}\tilde{m}_{cp'}}}{\frac{\sum_{1,C}^{c'}\tilde{m}_{c'p}}{\sum_{1,P}^{p'}\sum_{1,C}^{c'}\tilde{m}_{c'p'}}}
\end{equation}

Finally we define the entries of the biadjacency matrix $M$ of the undirected bipartite network analyzed in this work as:
\begin{equation}
\begin{cases}
\label{eq_rca}
    m_{cp}=1 & \text{when } \mbox{RCA}_{cp}\geq 1\\
    m_{cp}=0 & \text{otherwise}
\end{cases}
\end{equation}

This indicates that the connection (country-product link) is established if and only if the relative RCA is relevant (over the threshold), otherwise it can be ignored. Each row of $M$ represents the export basket of a given country (or its diversification  $k_c$), while each column represents the subset of producers of a given product (or its ubiquity $k_p$)\cite{hausmann2010country}.

\begin{equation}
	k_c=\sum_p m_{cp}\quad k_p=\sum_c m_{cp}
\end{equation}

\subsubsection*{Urbanization}

The data for the urban population from $1995$ to $2010$ are available at the World Bank database (\href{https://data.worldbank.org/}{https://data.worldbank.org/} Date of access: 16 July 2019).

\subsection*{Fitness and Complexity}

Fitness and Complexity are a metric for countries and products applied to bipartite binary matrix $M$ of the WTW \cite{Hidalgo:2007aa, Hidalgo:2009aa,tacchella2012new, tacchella2013economic,cristelli2013measuring}.
The basic idea of EC is to define a non-linear map through an iterative  process which couples the Fitness of countries to the Complexity of products. At every step of the iteration, the Fitness $F_c$ of a given country $c$ is proportional to the sum of the exported products, weighted by their complexity parameter $Q_p$.
 In particular, the Fitness $F_c$ for the generic country $c$ and Quality $Q_p$ for the generic product $p$ at the $n-$th step of iteration, are defined as:
\begin{equation}\label{eq:FFQQ}
\left\{
\begin{array}{c}
\tilde{F}^{(n)}_c=\sum_p m_{cp} Q^{(n-1)}_p\\
\\
\tilde{Q}^{(n)}_p=\dfrac{1}{\sum_c m_{cp} \frac{1}{F^{(n-1)}_c}}
\end{array}
\right.
\rightarrow
\left\{
\begin{array}{c}
F^{(n)}_c=\dfrac{\tilde{F}^{(n)}_c}{\langle \tilde{F}^{(n)}_c\rangle}\\
\\
Q^{(n)}_p=\dfrac{\tilde{Q}^{(n)}_p}{\langle \tilde{Q}^{(n)}_p\rangle}
\end{array}
\right.,
\end{equation}
where the symbols $\langle\cdot\rangle$ indicate the average taken over the proper set. The initial condition are taken as $F_c^0=Q_p^0=1\,\,\forall c\in N_c,\,\forall p\in N_p$, where $N_c$ and $N_p$ are the number respectively of countries and products 
(the convergence of the algorithm described by Eqs.(\ref{eq:FFQQ}) depends on the shape of the matrix $M$, as it has been discussed in \cite{Pugliese_2016}).

\subsection*{Bipartite Configuration Model (BICM)}

The Bipartite Configuration Model (BICM), as defined by \cite{saracco2015randomizing,saracco2016detecting}, is a null model of general applicability that is able to generate a grandcanonical ensemble of bipartite, undirected, binary networks in which the two layers Country and Products have respectively $C$ and $P$ nodes.
The ensemble generate by the BICM constrained the number of connections for each node, on both layers (in our case $d_c$ and $u_p$) to match, on average, the observed one.
Each network $\mathbf{M}$ in such ensemble is assigned a probability coefficient:
 
\begin{equation}
P(\mathbf{M}|\vec{x}, \vec{y})=\prod_cx_c^{d_c(\mathbf{M})}\prod_py_p^{u_p(\mathbf{M})}\prod_{c, p}(1+x_cy_p)^{-1},
\end{equation}

$x_c$ and $y_s$ are the Lagrange multipliers associated to the constrained degrees.

Constraining the ensemble average values of countries and products degree induces the probability that a link exists between country $c$ and industry sector $p$ independently of the other links:

\begin{equation}
p_{cp}=\frac{x_cy_p}{1+x_cy_p}.
\label{prob}
\end{equation}

The numerical values of the unknown parameters $\vec{x}$ and $\vec{y}$ have to be determined by solving the following system of $C+P$ equations, which constrains the ensemble average values of countries diversification and products ubiquities to match the real values, $\langle d_c\rangle=d_c^*,\:c=1\dots C$ and $\langle u_p\rangle=u_p^*,\:p=1\dots P$.

 Where $\{d_c^*\}_{c=1}^C$ and $\{u_p^*\}_{p=1}^S$ are the real degree sequence of countries, and industry sectors respectively, and $\langle \cdot\rangle$ represents the ensemble average of a given quantity, over the ensemble measure defined by Eq.\ref{prob} - as $\langle d_c\rangle=\sum_sp_{cp}$ and $\langle u_s\rangle=\sum_cp_{cp}$. Indicated with an asterisk, ``$\ast$" are the parameters that satisfy the systems.

\subsubsection*{Similarity Motifs}

In the present work we have sampled the grand canonical ensemble of binary, undirected, bipartite networks induced by the BiCM, according to the probability coefficients $P(\mathbf{M}|\vec{x}^*, \vec{y}^*)$ and calculated the average and variance of the motif $\mu_{\mbox{sim}}$, define as b-motif6 in \cite{simmons2019bmotif}.

The Similarity Motif represents the symmetric and complete connections between two countries $c,c'$ and two industry sectors $p,p'$. The number of similarity motifs is:

\begin{equation}
	\mu_{\mbox{sim}}=\frac{1}{4}\sum_{c=1}^C\sum_{c=1}^C\mathcal{Z}_{cc'}(\mathcal{Z}_{cc'}-1)-\frac{1}{4}\sum_{c=1}^C d_c(d_c-1)
	\label{eq_sim}
\end{equation}

with $\mathcal{Z}$ is the matrix of dimension $(C,C)$, that represents the projection of $M$ over the countries. Each entry $\mathcal{Z}_{cc'}$ counts the number of industry sectors in common between the countries $c$ and $c'$, it is defined as: $\mathcal{Z}_{cc'}=\sum_{s=1}^S M_{cs}M_{c's}=MM^T$

This motif represents the co-occurrence of two products in two countries' export basket within the bipatite matrix of the country exports. 
The accuracy of the BiCM prediction in reproducing the value of quantity $\mu_{sim}$ please follows \cite{saracco2015randomizing}.

\begin{figure}[!h]
\centering\includegraphics[width=0.99\textwidth]{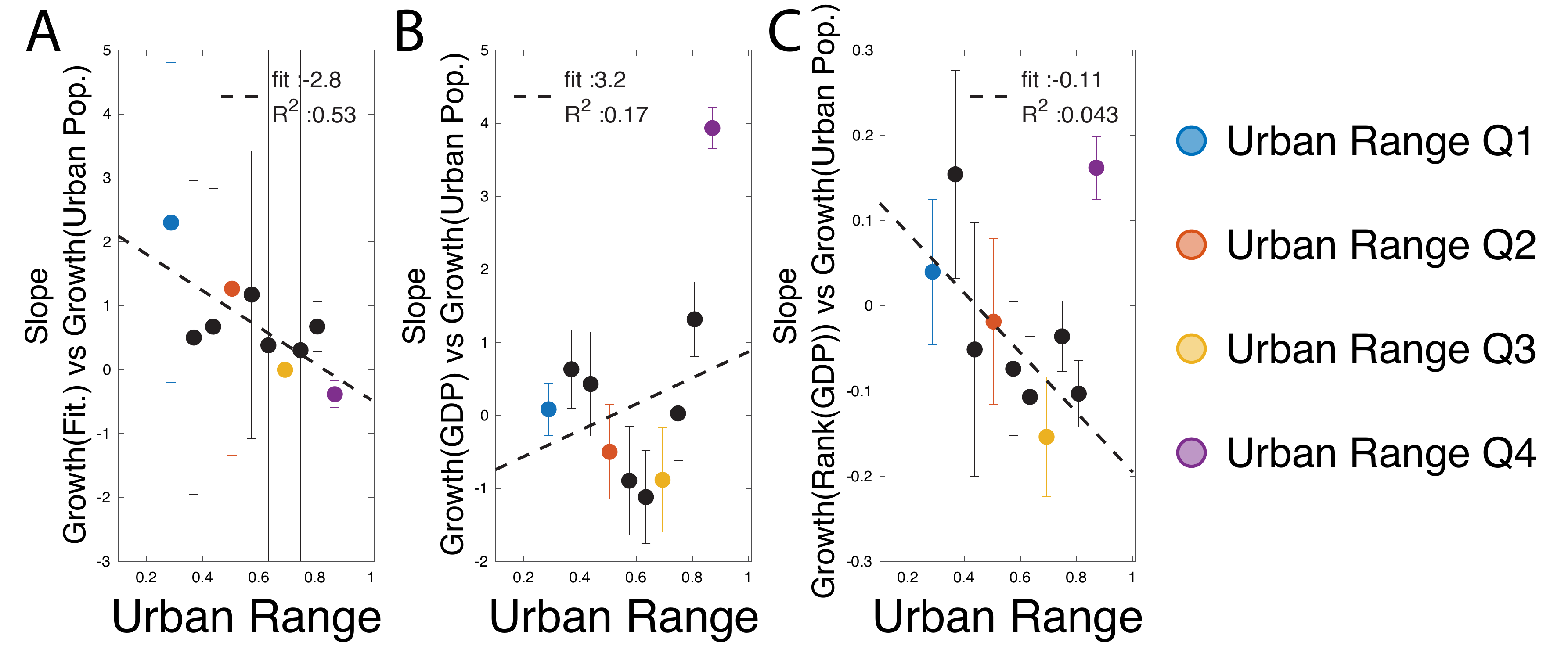}

\caption{
Slope coefficient of a sliding window across $25\%$ of the countries (corresponding to 36 countries) of respectively its Fitness Growth Rate \textbf{A} - GDP Growth Rate \textbf{B} - GDP Ranking Growth Rate \textbf{C}  vs. Urban Population Growth Rate. The error bar corresponds to the fit's  $95\%$ confidence interval. The colors follow the Urban Range scheme.}
\label{fig_5}
\end{figure}

\subsection*{Urban Range vs Fitness and GDP\label{Urban}}

To validate our analysis of the relation between the country's fitness and the urbanization process we analyzed the urbanization growth rate in relation to the growth rate of three different metrics:  the country Fitness (\ref{fig_5}A), country GDP (\ref{fig_5}B), and country GDP ranking (\ref{fig_5}C)  between 1995 and 2010. 

We study the variation of the slope coefficient of a sliding window across $20\%$ of the countries urban range and the three metrics above.
Both the metrics extracted from the GDP do not have statistical significant results.
Although the growth rate of fitness in relation with the urbanization growth rate manifest a linear relation  (Fig. \ref{fig_5}A) with an $R^2=0.53$ as \ref{fig_3}B, we notice that the fitness ranking is a more reliable tool than the raw fitness value\cite{pugliese2017complex}. The fitness ranking provides a more stable dynamics across each sliding window.

\section*{Acknowledgements}

Riccardo Di Clemente as Newton International Fellow of the Royal Society acknowledges support from the Royal Society, the British Academy, and the Academy of Medical Sciences (Newton International Fellowship, NF170505). We authors would like to thank Fabio Saracco, Enrico Ubaldi, Bernardo Monechi, Andrea Zaccaria, Andrea Gabrielli, Luciano Pietronero and Marta C. Gonz\'alez for the insightful discussions and comments.

\section*{Author contributions statement}
R.D.C. and E.S. designed the study and performed the research. R.D.C. analyzed the data a generated the plots, E.S. created the Map. R.D.C., E.S. and M.B. wrote the paper. All the authors gave the final approval for the publication.

\section*{Additional information}

\textbf{Competing interests} The authors declare no competing interest.

The corresponding author is responsible for submitting a \href{http://www.nature.com/srep/policies/index.html#competing}{competing interests statement} on behalf of all authors of the paper. This statement must be included in the submitted article file.

\end{document}